\documentclass[aps,prd,twocolumn,groupedaddress,nofootinbib]{revtex4-1}

\usepackage{amsfonts,amsmath,amssymb}
\usepackage{mathrsfs}
\usepackage{nicefrac}
\usepackage{graphicx}
\usepackage{color}
\usepackage{soul} 
\usepackage{hyperref}
\usepackage{accents}
\usepackage{subfigure}
\usepackage[normalem]{ulem}
\usepackage{hyperref}
\hypersetup{
    colorlinks=true, 
    linktoc=page,    
    linkcolor=blue,  
    urlcolor=magenta
}

\newcommand{\be}{\nopagebreak[3]\begin{equation}}
\newcommand{\ee}{\end{equation}}

\newcommand{\bee}{\nopagebreak[3]\begin{equation*}}
\newcommand{\eee}{\end{equation*}}

\newcommand{\ba}{\nopagebreak[3]\begin{eqnarray}}
\newcommand{\ea}{\end{eqnarray}}

\newcommand{\baa}{\nopagebreak[3]\begin{eqnarray*}}
\newcommand{\eaa}{\end{eqnarray*}}

\newcommand{\la}{\label}
\newcommand{\nn}{\nonumber}

\newcommand{\f}{\frac}

\begin{document}

\title{Black hole quantum atmosphere for freely falling observers}

\author{Ramit Dey}
\email[]{ramitdey@gmail.com}
\affiliation{Department of Theoretical Physics, Indian Association for the Cultivation of Science, 
Kolkata-700032, India}

\author{Stefano Liberati}
\email[]{liberati@sissa.it}
\affiliation{SISSA, 
Via Bonomea 265, 34136 Trieste, Italy and INFN, Sezione di Trieste;\\
IFPU - Institute for Fundamental Physics of the Universe, Via Beirut 2, 34014 Trieste, Italy}

\author{Zahra Mirzaiyan}
\email[]{Z.Mirzaiyan@gmail.com}
\affiliation{Erwin Schr\"{o}dinger International Institute for Mathematical Physics, University of Vienna, Vienna, Austria}
\affiliation{Isfahan University of Technology, Isfahan 84156-83111, Iran}
\author{Daniele Pranzetti}
\email[]{dpranzetti@perimeterinstitute.ca}
\affiliation{Perimeter Institute for Theoretical Physics,
31 Caroline St. N, N2L 2Y5, Waterloo ON, Canada}

\begin{abstract}
We analyze Hawking radiation as perceived by a freely-falling observer and try to draw an inference about the region of origin of the Hawking quanta. To do so, first we calculate  the energy density from the stress energy tensor, as perceived by a freely-falling observer.  
Then we compare this with the energy density computed from an effective temperature functional which  depends  on the state of the observer. The two ways of computing these quantities show a mismatch at the light ring outside the black hole horizon. To better understand this ambiguity, we show that even taking into account the (minor) breakdown of the adiabatic evolution of  the temperature functional which has a peak in the same region of the mismatch, is not enough to remove it.  We argue that the appearance of this discrepancy can be traced back to the process of particle creation by showing how the Wentzel--Kramers--Brillouin approximation for the field modes breaks down between the light ring at $3M$ and $4M$, with a peak at $r=3.3M$ exactly where the energy density mismatch  is maximized.
We hence conclude that these facts strongly support a scenario where the Hawking flux does originate from a ``quantum atmosphere" located well outside the black hole horizon. 
\end{abstract}

\maketitle
\section{Introduction}
By studying quantum fields in curved spacetime, Hawking \cite{hawking1975} showed that black holes must evaporate due to the semi-classical effect of particle creation.
Hawking radiation can be heuristically explained via various mechanisms, such as a tunnelling effect, where the Hawking pair gets separated across the black hole horizon. 

In this picture, the particle with positive energy tunnels out of the horizon while its partner with negative energy tunnels across it, causing a decrease in mass of the black hole. It can be shown that such a process leads a Planckian spectrum of  particles at the asymptotic future infinity \cite{Parikh:1999mf}. 

An alternative picture is based on the idea that the strong tidal forces near the black hole horizon prevent the re-annihilation of a particle and antiparticle pair formed spontaneously from the vacuum. This would be normally forbidden as energy conservation on a stationary spacetime would imply that one of the particles must be endowed with a negative energy equal and opposite in sign to that of its partner. 

However, the horizon of  a non-rotating black hole shields an ergoregion where negative energy states (with respect to asymptotic observers) are allowed. So the negative energy particle can go ``on-shell" just inside the black hole horizon, falling then into the black hole and reducing its ADM mass while its partner can go on-shell outside of the horizon and escape to the future asymptotic infinity region \cite{hawking1979general, Parker1977, Dey:2017yez}.

Moreover, in support of this second possibility, a  recent claim based on calculating the effective size of a radiating body \cite{Giddings:2015uzr} is that, indeed, the Hawking quanta originate from a near horizon region, referred to as the ``quantum atmosphere". 

In   \cite{Dey:2017yez}, some of the authors  of this manuscript supported this claim with two different arguments. The first was based on the gravitational analog of the Schwinger effect for particle production by the tidal force outside a black hole horizon, motivated by the just mentioned alternative heuristic picture for Hawking radiation; this was used  to relate the energy of the thermal spectrum at infinity to the radial position at which the outgoing partner goes on-shell. 

The second argument made use of a full calculation of the renormalized stress energy tensor (RSET) in (1+1) dimensions  to derive the energy density for an observer at constant Kruskal position and investigate their behaviour in this quantum atmosphere outside the horizon. Both arguments provide evidence that most of the contribution to the spectrum of radiation at infinity originates from a region around  $r\approx 4 M$ outside the black hole horizon.

Shedding light on  the location where (most of) the Hawking quanta originate from can have important physical implications. 
First of all,  identifying the region of particle creation at finite distance from the horizon would provide an argument to naturally circumvent the transplanckian issue \cite{Jacobson:1991gr}. Moreover, it could alleviate the well known firewall problem \cite{Almheiri:2012rt} for those proposals  relying on correlations between the late and the early time radiation fluxes in order to retrieve the information at null infinity (this process to is sometimes referred to as
``purification'' of the late Hawking
radiation).

In fact, if the information content of the Bekenstein--Hawking entropy formula gets encoded in the entanglement structure of outgoing modes created at finite distance from the horizon, the breaking of entanglement across the horizon required by the purification process would generate a divergent energy density at the horizon encountered by the infalling observer with a much lower energy scale. Removal of this ``firewall'' through dynamical effects extended at this new finite distance would clearly require some non-local physical mechanisms of the kind advocated for in
\cite{Giddings:2012bm, Giddings:2012gc}.

Therefore, the origin of Hawking quanta represents  a physically very relevant question.
In this paper, we want to investigate and understand better this issue from a different perspective. More precisely, the second argument given in  \cite{Dey:2017yez} relied on the shape of the energy density and the flux of a test scalar field as perceived by an observer free falling {\it in Kruskal coordinates}. While such an observer has zero acceleration both at the horizon and at infinity, this is not true at finite distance from the horizon. Then, one might worry that the peak in energy density and flux found at around $4 M$ for such an observer is due to the non-zero acceleration in that region and it could be traced back to the Unruh effect contribution to the perceived radiation, rather than a real particle creation process. 

In order to further investigate  the physical nature and behavior of Hawking particle creation in the quantum atmosphere outside the horizon, we shall  consider here the physical information provided by the energy density as perceived by a truly freely-falling observer. Our first task is thus to derive these quantities for the Unruh vacuum ({as this is the physically relevant one to study the evaporation of non-eternal black holes soon after their formation~\cite{Barcelo:2007yk}}) in the Painlev\'{e}--Gullstrand coordinates, which are naturally adapted  to a freely-falling observer. {Given that the RSET is analytically known only in $1+1$ dimensions we shall then work in this setting throughout this paper.}

Not surprisingly, we recover  in Section \ref{sec:RSET}  results   well known in the literature through different methods 
\cite{Candelas:1980zt, Greenwood:2008zg, Barbado:2011dx, Eune:2014eka}. Namely, both the energy density and the energy flux are monotonically growing from zero at infinity to a finite value at the horizon. This monotonic behavior, however, does not necessarily contradict the argument of   \cite{Dey:2017yez}  based on a peak located at finite distance from the horizon for a Kruskal freely-falling observer. 

Indeed, it has been argued in \cite{Barbado:2011dx} that the final increase of effective temperature for  freely-falling observers is due to their non-zero radial velocity and it can be accounted for by a  Doppler shift factor with respect to a  stationary observer with zero radial velocity. Given that this  Doppler blue-shift factor diverges at the horizon, it is plausible to expect that a possible peak in energy density as measured by a freely-falling observer with non-zero radial velocity gets washed out by this kinematical effect, hiding the truly physical location of particle creation.

In Section \ref{sec:ET} we derive the effective temperature introduced in~\cite{Barcelo:2010pj, Barbado:2011dx, Barbado:2012pt} for our choice of freely-falling observer. If the Hawking flux is all the way to the horizon thermal we do expect such an effective description to match the one obtained  from the RSET. To further strengthen the case we check the viability of the condition for an adiabatic evolution of the perceived temperature for a freely falling observer.

In Section \ref{sec:disc} we then contrast the energy density  computed from the RSET in Section \ref{sec:RSET} with that obtained from the effective temperature method for the same kind of freely-falling observer. We find that the two energy densities present a minor discrepancy in the region $r\approx 3M\sim4 M$, with a peak overlapping to the one of the effective temperature adiabaticity violation. 
We thus take into account the non-adiabaticity of the temperature functional and include its effect in the energy density to compare it with the energy density obtained from the RSET. However, the net effect is a decreased accuracy of the effective temperature method in reproducing the RSET results.

Therefore, this mismatch seems to signal the location where a physical process is taking place, suggesting an origin of Hawking quanta at finite distance outside the horizon.
To further support our interpretation,
we study  in Section \ref{sec:Ad} the adiabaticity of the test field modes. This allows us to test where the Wentzel--Kramers--Brillouin (WKB) approximation breaks down, which  is often taken as an indicator of a particle creation process. We find a peak in the violation condition for the field modes around  $r=3.3 M$, in stark agreement to the location of the peak
of the discrepancy between the RSET energy density and
the effective temperature one.

A final discussion of these results is presented in Section \ref{sec:discus}.


 \section{Calculation of RSET in Painlev\'{e}--Gullstrand coordinate}  \la{sec:RSET}

 An accurate way to probe Hawking radiation as perceived by an observer is by studying the stress energy tensor of a test field. As shown in \cite{Dey:2017yez}, an observer falling 
 with small acceleration outside the horizon\footnote {Often referred to as a {\it Kruskal observer}, it has zero acceleration and radial velocity at the horizon crossing but its not a strictly geodesic observer globally.}
measures a peak of the energy density and flux at $4.3M$, hinting at the presence of a quantum atmosphere outside the black hole horizon. 
 
 In this section we are interested in computing the energy density and flux as measured by an observer freely falling into the black hole in a coordinate system adapted to geodesic motion. While such an observer has zero acceleration throughout its trajectory, it will have a non-zero radial velocity at horizon crossing and a non-zero acceleration with respect to infinity, which might give rise to some kinematical effects \cite{Barbado:2012pt}. To study the behavior of the energy density and flux as perceived by this observer we first setup the Painlev\'{e}--Gullstrand coordinates and then obtain the RSET components in this coordinate system using a coordinate transformation. 
 
Given the black hole metric in $1+1$ dimensions in Schwarzschild  coordinates
\begin{equation}\label{Schw metric}
ds^2=\left(1-\frac{2M}{r}\right)dt^2-
\frac{dr^2}{\left(1-\frac{2M}{r}\right)}\,,
\end{equation}
where $M$ is the mass of the black hole, we can introduce the Painlev\'{e}--Gullstrand coordinates through the  definition of a new time coordinate as 
\begin{equation}\label{Painleve time}
t_{p}=t-f(r)
\end{equation}
for some arbitary function $f(r)$, such that
\begin{equation}\label{fprime}
f^{\prime} (r)=-\frac{1}{1-\frac{2M}{r}}\sqrt{\frac{2M}{r}}\,.
\end{equation}
Substituting Eq.(\ref{fprime}) in Schwarzschild metric Eq.(\ref{Schw metric}) one gets
\begin{equation}\label{painleve metric}
ds^2=\left(1-\frac{2M}{r}\right)dt_p^2-2\sqrt{\frac{2M}{r}} dt_{p} dr-dr^2\,.
\end{equation}
We see the spatial slices of the metric (\ref{painleve metric}) corresponds to the flat metric in spherical polar coordinates. Also, there is no coordinate singularity at the Schwarzschild radius $(r=2M)$. The time coordinate of the Painlev\'{e}--Gullstrand metric is the same as the proper time of a freely-falling observer who starts from infinity at zero velocity.

We denote the Painlev\'{e}--Gullstrand coordinates as $(t_{p},x)$ and the Schwarzschild coordinates as $(t,r)$, with $r=x$. The Jacobian for the coordinate transformation is given by

\[
\frac{\partial (t_{p},x)}{\partial(t,r)}=
\left({\begin{array}{cc}
\frac{{\partial t_{p}}}{{\partial t}}&\frac{{\partial t_{p}}}{{\partial r}}\\
\frac{{\partial x}}{{\partial t}}&\frac{{\partial x}}{{\partial r}}\\
\end{array}}\right)
=\left({\begin{array}{cc}
1&-f^{\prime}\\
0&1\\
\end{array}}\right)\,,
\]
and the inverse of the transformation matrix is given by
\[
\frac{\partial (t,r)}{\partial(t_{p},x)}=
\left({\begin{array}{cc}
\frac{{\partial t}}{{\partial t_{p}}}&\frac{{\partial t}}{{\partial x}}\\
\frac{{\partial r}}{{\partial t_{p}}}&\frac{{\partial r}}{{\partial x}}\\
\end{array}}\right)
=\left({\begin{array}{cc}
1&f^{\prime}\\
0&1\\
\end{array}}\right)\,.
\]

The RSET components of the Schwarzschild spacetime is well known for various vacuum states \cite{Davies:1976ei, Unruh:1977ga, Birrell:1982ix}. Choosing the Unruh vacuum state \footnote{Throughout the paper we considered the case of black hole collapse and worked with Unruh vacuum state. For this case, unlike \cite{Barbado:2016ger}, we do not need to consider the contribution of the effective temperature with respect to the ingoing null coordinate to the energy density.} for a freely-falling observer we can write the RSET components in Painlev\'{e}--Gullstrand coordinates by means of a coordinate transformation 
 \begin{equation}\label{painto schwRSET}
 T^{\text{GP}}_{\alpha\beta}=\frac{\partial x^{\mu}}{\partial x^{\alpha}}\frac{\partial x^{\nu}}{\partial x^{\beta}} T^{\text{Schw}} _{\mu\nu}\,.
 \end{equation}
 This yields
 \begin{eqnarray}\label{Ttptp}
 T_{{t_{p}}{t_{p}}}&=& T_{tt}\,,\\
 T_{{t_{p}}{x}}&=&f^{\prime}T_{tt}+T_{rt}= T_{{x}{t_{p}}}\,,\\
T_{{x}{x}}&=& {f^{\prime}}^{2}T_{tt}+2 f^{\prime} T_{t r}+T_{r r}\,,
 \end{eqnarray}
where the components of RSET in terms of the Schwarzschild coordinates in the Unruh vacuum \cite{Unruh:1977ga} are given by
\begin{eqnarray}\label{RSET Com}
T_{tt}&=&\frac{1}{24\pi}\left(\frac{7 M^2}{r^4}-\frac{4 M}{r^3}+\frac{1}{32 M^2}\right)\,,\\
T_{tr}&=&-\frac{1}{24\pi}\frac{1}{ (1-\frac{2M}{r})}{\frac{1}{32 M}}\,,\\
T_{rr}&=&-\frac{1}{24\pi}\frac{1}{(1-\frac{2M}{r})^2}\left(\frac{M^2}{r^4}-\frac{1}{32M}\right)\,.
\end{eqnarray}

For a freely-falling observer, the velocity in Painlev\'{e}--Gullstrand coordinates is given by
 \begin{equation}\label{free falling velocity}
 V^{a}=\left(1,-\sqrt{\frac{2M}{r}}\right).
 \end{equation}
 Using this velocity,  the associated energy density as measured by such an observer  can be calculated as
 \begin{equation}\label{energy density}
{ \varepsilon}_{\text{RSET}}=T_{ab} V^{a} V^{b}=T_{t_{p} t_{p}}-2\sqrt{\frac{2M}{r}} T_{t_{p} x}+\frac{2M}{r} T_{xx}\,.
 \end{equation}

\begin{figure}[h]
\includegraphics[width=8.3cm]{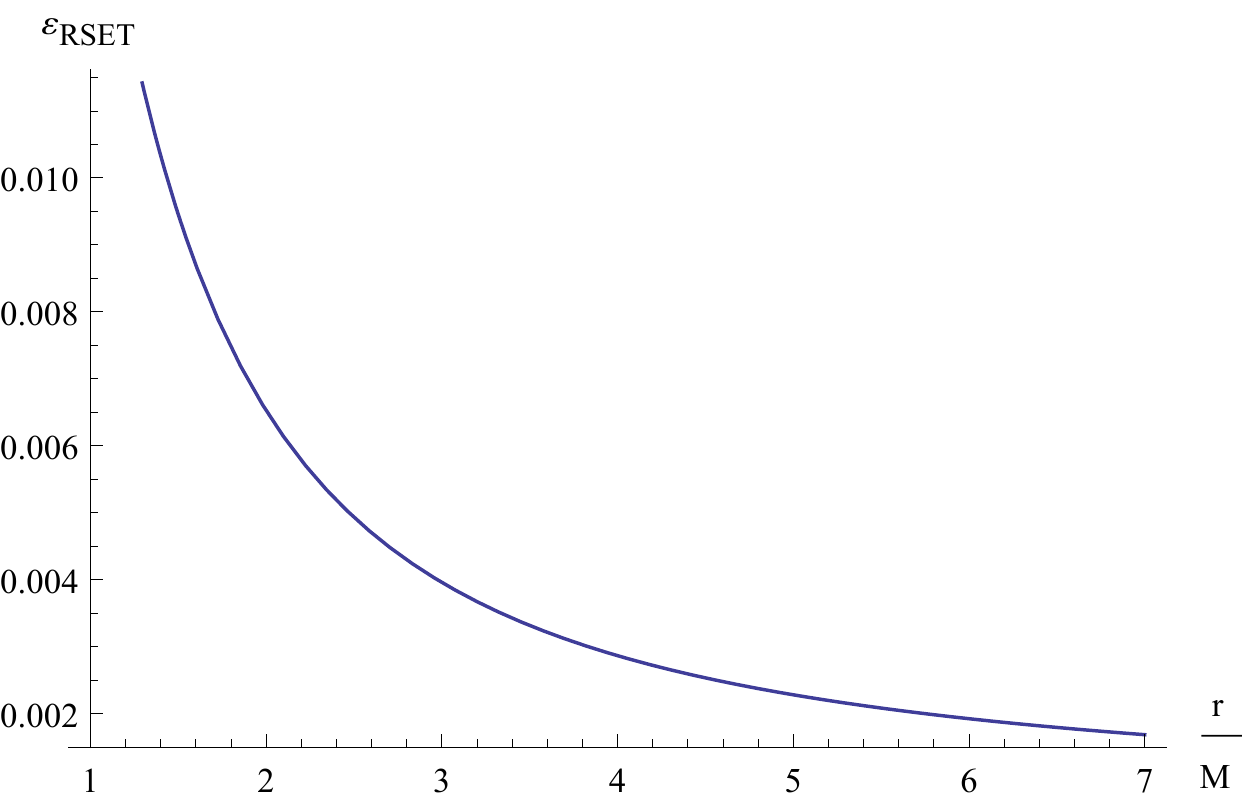}\\
 \caption[] {{Energy density as a function of distance $r$. As the observer get closer to the horizon the energy density increases but remains finite at the horizon.}}\label{figure:GT}
 \end{figure}

 In FIG. \ref{figure:GT} we plot the energy density as a function of the radial distance  $r$. We can see that the energy density increases as the observer gets closer to the black hole and eventually it remains finite while the observer crosses the horizon.
 
 We can also calculate the flux of energy as perceived by a freely-falling observer in Gullstrand-Painlev\'e coordinates. For calculating the flux we need the orthogonal vector with respect to the velocity.  We can use the othogonality condition to obtain this orthogonal vector as
\begin{eqnarray}\label{conditions}
g_{ab} n^{a} n^{b}&=&-1\,,\nonumber\\
n_{a} V^{a}&=&0\,,
\end{eqnarray}
 where the normal vector $n^a$ can be defined as 
 \begin{equation}\label{normal vector}
 n^{a}=(0,1)\,.
 \end{equation}
The flux in Painlev\'{e}--Gullstrand coordinates is then 
 \begin{equation} \label{flux}
 F_{\text{RSET}}=T_{ab} n^{a} V^{b}=-T_{t_{p} x}+\sqrt{\frac{2M}{r}} T_{xx}\,.
 \end{equation}
 
\begin{figure}[h]
\includegraphics[width=8.3cm]{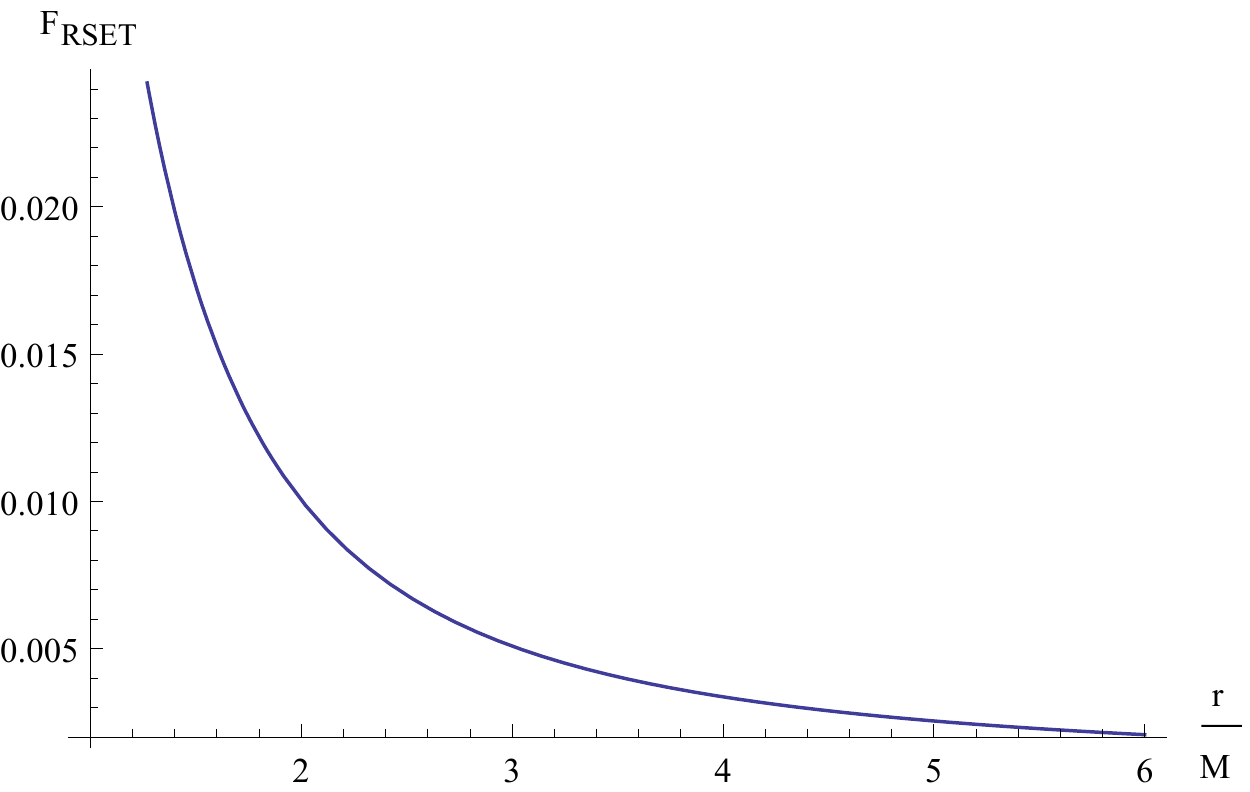}\\
 \caption[] {{Energy flux as a function of distance $r$. }}\label{figure:Flux}
 \end{figure}

 In FIG. \ref{figure:Flux} we plot this flux as a function of the radial coordinate $r$.
 
 Contrary to \cite{Dey:2017yez}, we do not see a peak in the energy density or the flux near the horizon. Hence, relying just on the RSET computation we cannot say much about the presence of the quantum atmosphere. As anticipated in the beginning, this monotonic behavior can be traced back to the non-zero radial velocity of the freely-falling observer considered here, which gives rise to kinematical effects. Therefore, we have to devise some way to extract physical information about the region of  origin of the Hawking quanta from the energy density and flux computed from the RSET.


\section{The effective temperature functional}\la{sec:ET}

To study how different observers would perceive Hawking radiation, an effective temperature function was introduced in \cite{Barcelo:2010pj, Barbado:2011dx,Barbado:2012pt}, which depends on the trajectory of the observer. We  now rederive this temperature functional for a freely-falling observer and study its adiabaticity, which will be later used to compute the perceived energy density to  compare with our findings in the previous section. 

\subsection{The effective temperature functional}
Considering a Schwarzschild black hole in $(t,r,\theta,\phi)$ coordinates, one can define an outgoing null coordinate as
\ba
\bar u=t-r^*\,,
\ea
where $r^*$ is the tortoise coordinate given by $r^*=r+2M \log({r/2M}-1)$. For an infalling observer, we can use its proper time to label different light rays it encounters while infalling. We can define the proper time associated with the observer as $\tau$ and then the timelike trajectory of the observer would be defined as $(t(\tau),r(\tau))$.
 The observer can define the labeling of the $\bar u= const$ rays that it encounters as  $U=U(\bar u)=\tau$ which gives us another set of null coordinates.
  
  For doing quantum field theory on a background containing such a black hole,  $U=U(\bar u)$ would determine the choice of the local vacuum state. We can define yet another set of null coordinates based on the proper time, $\tau$, of the observer (after proper synchronization) so that we can study the perception of the vacuum state related to $U=U(\bar u)$ for a specific observer. We define this null coordinate as
\ba
u=\tau-\tau_0\,,
 \ea 
 where $\tau_0$ is a constant used for synchronization.
 One can easily obtain the relation $U=U(u)$ using $\bar u=\bar u(u)$. Using this relation, it is possible to compute the Bogoliubov coefficients giving the particle content as perceived by the observer in the given vacuum state. One can define the effective temperature function as \cite{Barbado:2011dx,Barbado:2012pt}
\ba \la{temp}
\kappa_{\text{eff}}(u)=-\frac{d^2U}{du^2}/\frac{dU}{du}.
\ea
If $u$ corresponds to the future null coordinate for a Schwarzschild geometry, then $\kappa_{\text{eff}}$ contains information about the peeling of null geodesics and this would be the relevant quantity for calculating the Hawking temperature. In this case, the way $\kappa_{\text{eff}}$ is defined  contains information about  the peeling as well as about the vacuum state and the observer. 

After choosing an appropriate vacuum state by specifying $U(\bar u)$, \eqref{temp} can be written as
\ba
\kappa_{\text{eff}}&=&-\bigg( \f{d^2U}{d \bar u^2}/\f{dU}{d\bar u}\bigg)\f{d \bar u}{d u}-\f{d^2 \bar u}{d u^2}/\f{d \bar u}{d u}\nn\\
&=&\f{d \bar u}{d u}\kappa(\bar u)-\f{d^2 \bar u}{d u^2}/\f{d \bar u}{d u}\,,
\ea
where $\kappa (\bar u)$, defined as
\ba
\kappa (\bar u)=-\f{d^2U}{d\bar u^2}/\f{dU}{d \bar u}\,,
\ea
 is the state dependent temperature,  as it solely depends on the choice of the vacuum state and not on the trajectory of the observer.

One can check the value of $\kappa(\bar u)$ for different vacuum states by considering the corresponding $U(\bar u)$ relation. For the Unruh vacuum $U(\bar u)=-4Me^{-\bar u}/(4M)$, which gives $\kappa(\bar u)=\frac{1}{4M}$; similarly, for the Boulware vacuum $U(\bar u)=\bar u$, which gives $\kappa (\bar u)=0$.

As shown in \cite{Barbado:2012pt}, the effective temperature in the Unruh vacuum for an arbitrary observer having proper time $u$ and following a trajectory $(t(u),r(u))$,  is given by 
\ba \label{eff_temp}
\kappa_{\text{eff}}(u)=\sqrt{\f{1-v_l}{1+v_l}}\frac{1}{\sqrt{1-\f{2M}{r}}}\bigg( \kappa(\bar u)-\f{M}{r^2}\bigg)+a_p\,,
\ea
where $a_p$ is the proper acceleration of the observer and $v_l$ is the velocity of the observer with respect to the black hole as measured in a local inertial frame, which is given as 
\begin{equation}
v_{l}=\frac{V_r}{\sqrt{1-\frac{2M}{r}+V_r^2}}\,,
\end{equation}
where $V_r=\sqrt{{2M}/{r}}$ is the radial velocity of the freely-falling observer.

\subsection{Adiabatic evolution of the Temperature functional}

\begin{figure}[h]
\includegraphics[width=8.3cm]{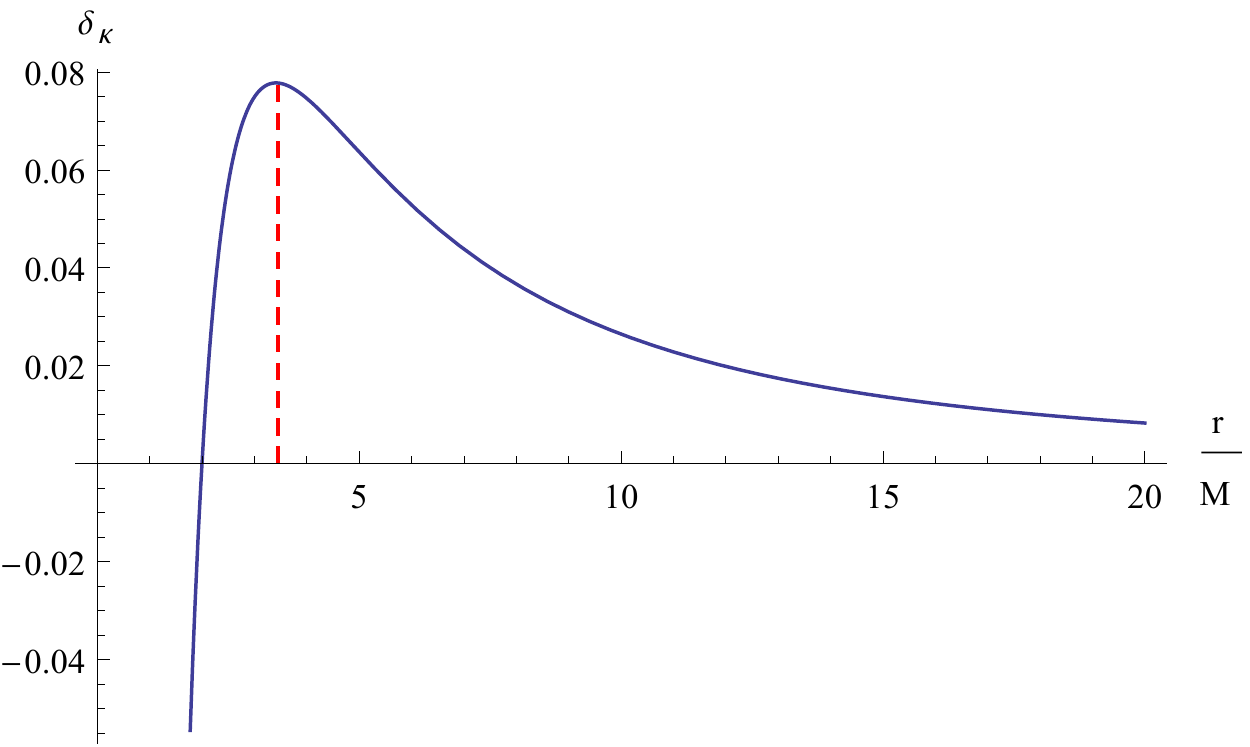}\\
 \caption[] {{Adiabaticity of the temperature functional as a function of distance $r$. The peak  at $r=3.4M$ in the plot shows where the maximum violation of the adiabatic condition happens.}}\label{figure:add}
 \end{figure}

To verify if this temperature function is an accurate estimator of the effective temperature perceived by an observer all along the trajectory as it falls into the black hole,  we do check that the perceived variation in time of the effective temperature functional for the freely-falling observer always satisfy an adiabatic evolution condition. This adiabatic condition of the effective temperature is estimated by
\begin{equation}\label{adiabatic}
\delta_\kappa=\frac{\dot{\kappa}_{\text{eff}}}{{\kappa_{\text{eff}}}^2}\ll 1\,,
\end{equation}
where, from \eqref{eff_temp}, we obtain  $\kappa_{\text{eff}}$ for a freely-falling observer in Unruh vacuum by setting $a_p=0$, $\kappa(\bar u)=\frac{1}{4M}$ and $v_l=-\sqrt{2M/r}$, as

\begin{equation} \label{keff}
\kappa_{\text{eff}}=\frac{1}{\left(1-\sqrt{\frac{2M}{r}}\right)}\bigg(\frac{1}{4M}-\frac{M}{r^2}\bigg)\,.
\end{equation}
To know where the maximum violation of adiabatic condition (\ref{adiabatic}) happens, we  plot the function $\delta_\kappa$ with respect to the radial distance $r$ in FIG. \ref{figure:add}.  

 As we can see from the plot,  the adiabatic condition is not strictly violated anywhere; however, there is a peak at  $3.4 M$ around which {its validity is minimized}. We will come back to this feature in a moment.
 



 \section{discrepancy in energy density}\la{sec:disc}

From the above discussion it is clear that the effective temperature functional allows to define a perceived energy density by any observer as the one associated to a thermal bath at the observer dependent temperature set by $\kappa_{\rm eff}$.
On the other hand, the RSET computed in Section \ref{sec:RSET} gives an accurate estimate of the energy density as measured by any observer. In this section we try to quantify the difference between the energy density obtained using the RSET  and the same quantities obtained from the effective temperature for freely falling observers and see if an eventual discrepancy can be related to the violation of adiabaticity.

\begin{figure}[h]
\includegraphics[width=8.3cm]{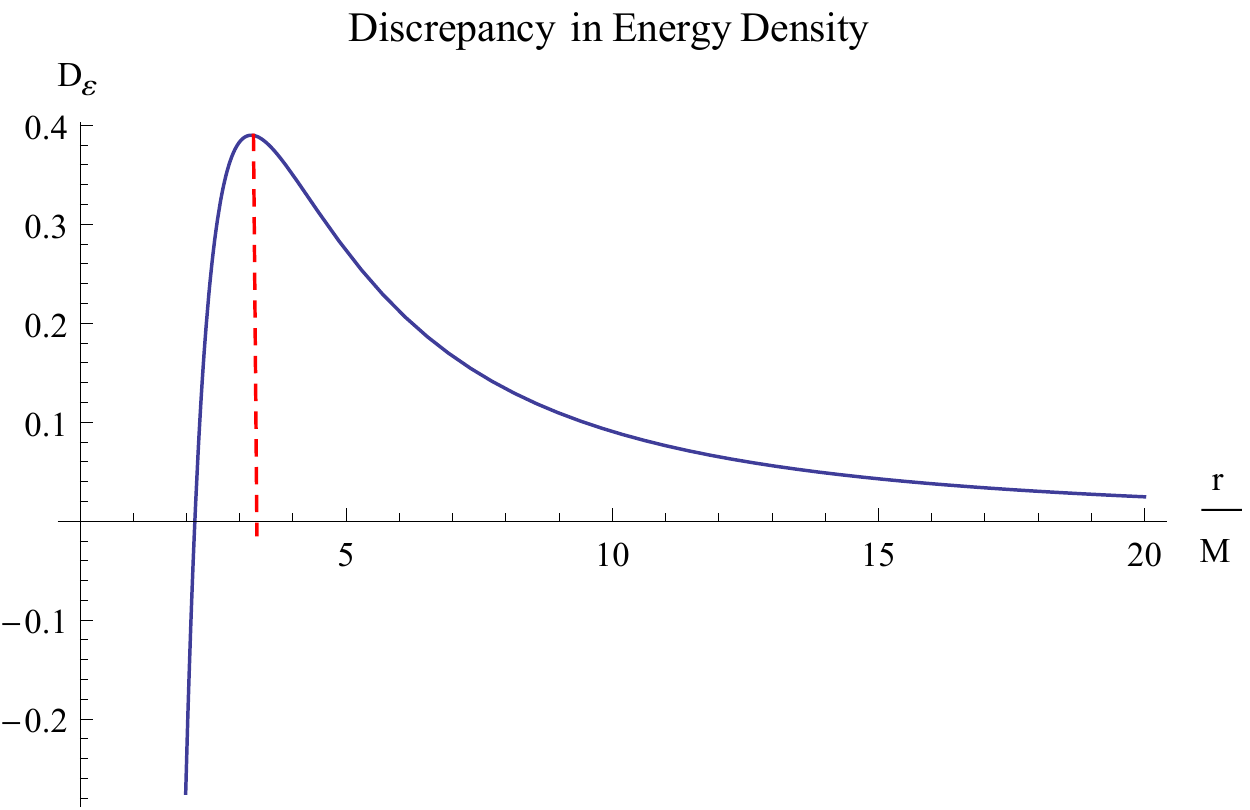}\\
 \caption[] {{Discrepancy in energy density calculated by two methods where there is a peak at ($r=3.2M$).}}\label{figure:dis}
 \end{figure}

 If we assume a Planckian spectrum for the radiation, we can use the effective temperature $\kappa_{\text{eff}}$  \eqref{keff} to  compute the energy density as 
\begin{equation}\label{ro}
{\varepsilon_{{\rm eff}}}=\frac{1}{48 \pi}(\kappa_{\text{eff}})^2\,.
\end{equation}
To see how this energy density differs from the energy density obtained in \eqref{energy density}, we  plot the difference between the two in FIG. \ref{figure:dis}, by defining the discrepancy in energy density as
\be
D_{\varepsilon}=\frac{1}{\varepsilon_{\text{infinity}}} (\varepsilon_{RSET}-\varepsilon_{{\rm eff}})\,,
\ee
where by $\varepsilon_{\text{infinity}}$ we mean the energy density at infinity, namely 
\be
\varepsilon_{\text{infinity}}=\frac{\kappa_{\text{infinity}}^2}{48\pi}= \frac{1}{48\pi} \frac{1}{(4M)^2}\,.
\ee
%

From the plots we see that the  energy density  calculated in \eqref{energy density} for a freely-falling observer is consistent with the one derived via the effective temperature method at infinity, but there is a small discrepancy as one moves closer to the horizon, reaching a peak at finite distance before vanishing again at horizon crossing. Notice that, in general, such a good agreement in most of the exterior region is not an obvious expectation as the RSET encodes physical information which is insensitive to the acceleration of the observer; on the other hand, the effective temperature function was introduced exactly to analyze the perception by different observers in various vacua and it contains information that is sensitive to the observer's acceleration. However, since our choice of observer has zero proper acceleration, it is natural to expect that the two approaches should capture the same physical content of the emission process. We thus believe that the peak in the discrepancy shown in FIG. \ref{figure:dis}, while not necessarily alarming, should not be overlooked, as investigation into it might reveal important insights into the physical nature of departure between the two methods.

One reason for this discrepency could be based on the fact that the assumption of the Plankian spectrum throughout the exterior of the black hole is not correct as the temperature function shows departure from adiabaticity in the same region outside the horizon. 

\begin{figure}
\includegraphics[width=8.3cm]{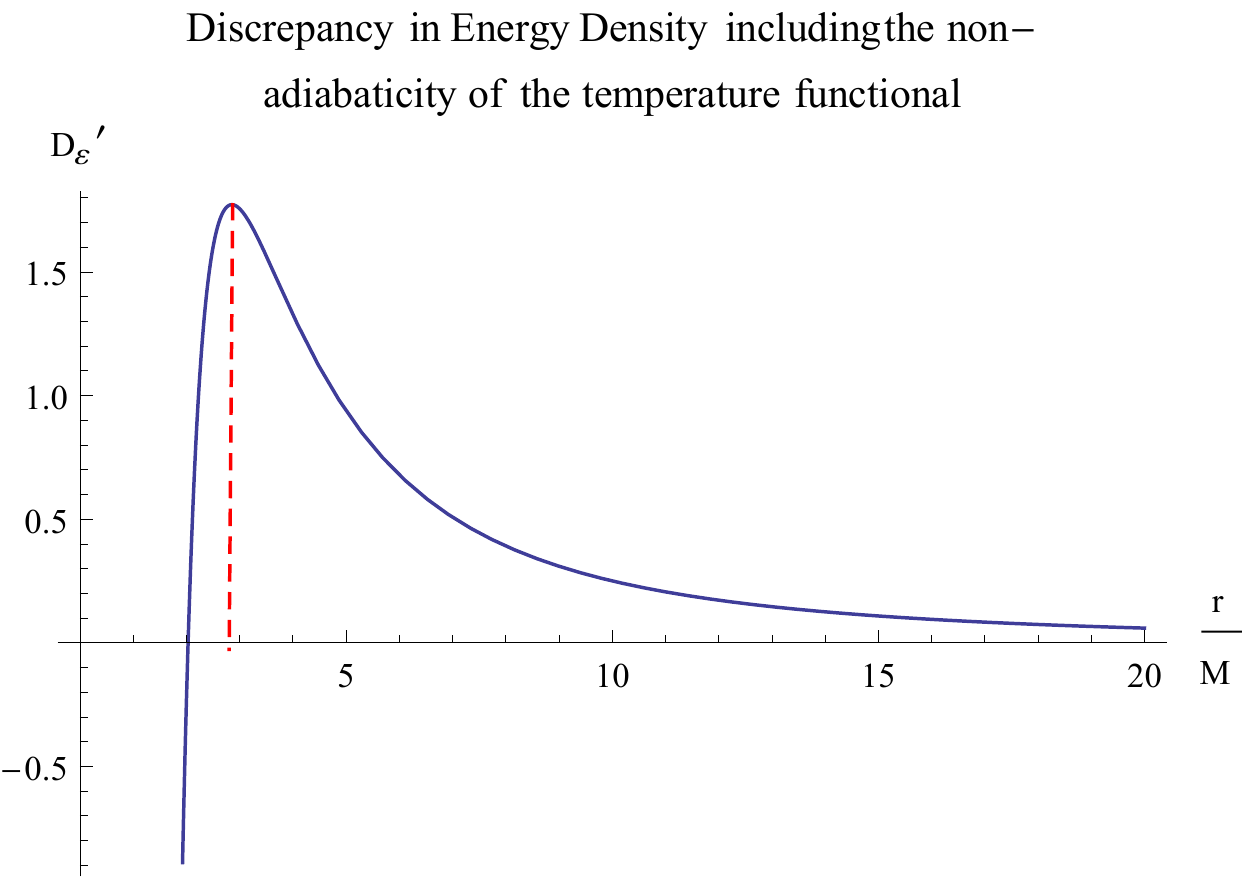}\\
 \caption[] {{Discrepancy in energy density calculated by two methods, after inclusion of the effects due to the non-adiabaticity of the temperature functional, where there is a peak at ($r=2.9M$).}}\label{figure:dis2}
 \end{figure}

This becomes more evident by comparing the plots in FIG. \ref{figure:add} and   FIG. \ref{figure:dis}, as the non-vanishing difference between the two energy densities overlaps with the region where the adiabaticity of the temperature function is not strict. We can try to capture the effect of the non-adiabiticy of the temperature functional by including a term that is given by the rate of change of the temperature functional with respect to the proper time in \eqref{ro}. Doing so, Eq.\eqref{ro} gets modified as \cite{Barbado:2012pt}

\begin{eqnarray} \label{energy_mod}
{\varepsilon'_{{\rm eff}}}=\frac{1}{24 \pi}\bigg[\frac{1}{2}\kappa_{\text{eff}}^2+\frac{d \kappa_{\text{eff}}}{d \tau}\bigg].
\end{eqnarray}
 Now, plotting the difference of this energy density with the energy density obtained from the RSET, i.e.
 \be
D'_{\varepsilon}=\frac{1}{\varepsilon_{\text{infinity}}}(\varepsilon_{RSET}-\varepsilon'_{{\rm eff}}) ,
\ee
we see that the  discrepancy is still there FIG. \ref{figure:dis2}, infact it increased even more with the peak just slightly shifting more towards the horizon.
 This indicates that the near violation of the adiabatic condition of the temperature functional is not causing this discrepancy,
 leaving as logical alternative the location of the
 particle creation origin of the Hawking quanta in that region. To probe this, finally, we check wether  the WKB approximation breaks down in the same region.

\section{Adiabatic condition}\la{sec:Ad}

To support the interpretation we gave above about the discrepancy between  the energy densities computed from the two different methods as associated to the location where physical particle creation is taking place, one can directly compute the adiabaticity of the field modes to verify if the WKB approximation breaks down in this region, as this is an indicator of
particle creation\footnote{See for example \cite{Jacobson:2003vx},  but also \cite {Schutzhold:2013mba}
for an alternative point of view in the case of dispersive media.}.

In order to provide this further check, we first need to know the frequency of the modes as measured by a freely-falling observer with respect to its proper time.
Let us consider again the Schwarzschild metric \eqref{Schw metric} and denote $f(r)=1-{2M}/{r}$.
%
The radial velocity for the freely-falling observer is given by $V_r=\sqrt{{2M}/{r}}$ 
and the four-momentum of the outgoing particle (mode) as measured by a freely-falling observer along the null ray $u$=constant reads 
\begin{eqnarray}\label{kafree}
k^a=\omega_{0} \left(\frac{1}{f(r)},1\right)\,,
\end{eqnarray}
where $\omega_0$ is a constant
which we can think of as $\omega_{Hawking}$ at $r \to \infty$. For the free-fall frame the velocity tangent to the path of an ingoing
(timelike) geodesic is given by
 \begin{eqnarray}\label{velotan}
 \beta ^{a}=(1,-V_r f(r))
 \end{eqnarray}
 and   the frequency can be defined as 
\begin{eqnarray}\label{frequ}
\omega=\frac{\beta^{a} k_{a}}{\parallel\beta\parallel}=\frac{\omega_{0}}{1-V_r}\,.
\end{eqnarray}
The rate of change of the frequency with respect to the Schwarzschild time, $\dot{\omega}$, is then
\begin{eqnarray}\label{omegdot}
\dot{\omega}=\frac{d\omega}{dt}=\frac{d\omega}{dV_r}\frac{dV_r}{dr}\frac{dr}{dt}=\frac{\omega_{0} V_r^3 (1-V_r^2)}{(1-V_r)^2}\,.
\end{eqnarray}

\begin{figure} [h]
\includegraphics[width=8.3cm]{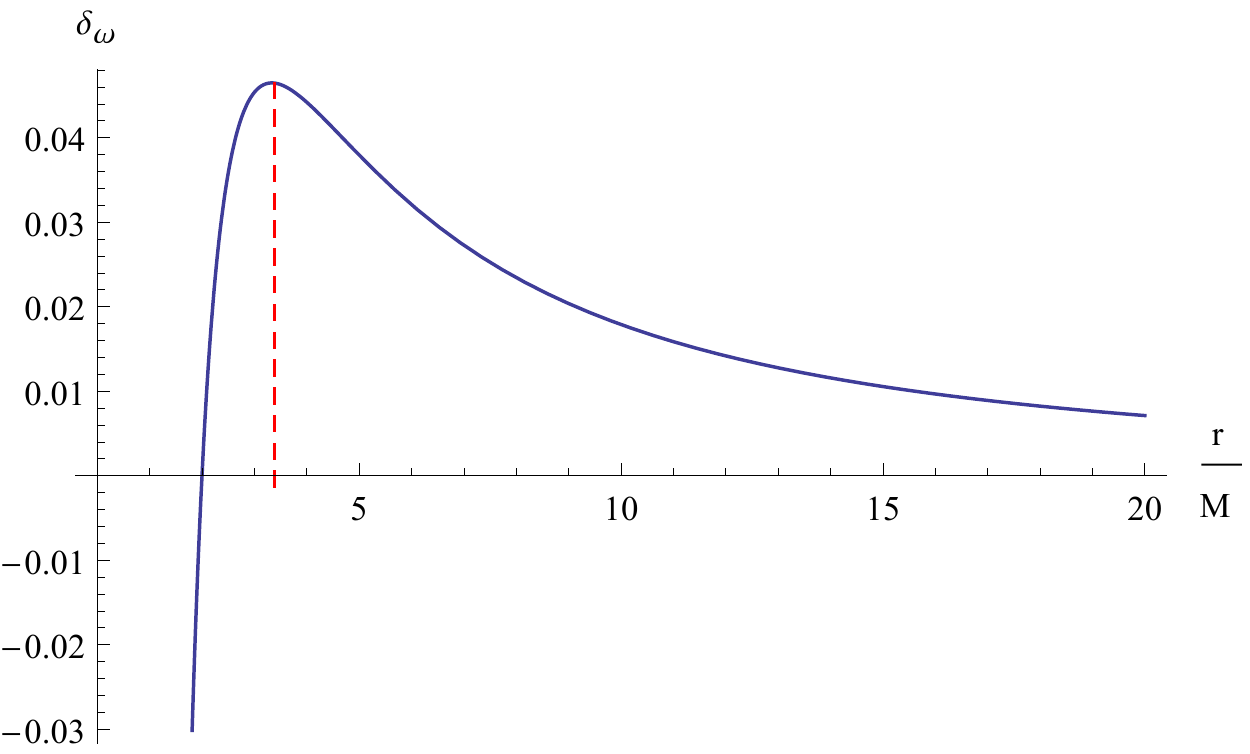}\\
 \caption[] {{$\delta_{\omega}$ as a function of distance $r$. The peak in the plot at $r=3.3 M$ shows where the maximum violation of the adiabatic condition for the field modes takes place.}}\label{figure:add3}
 \end{figure}

The adiabaticity of the modes is defined by the condition $\frac{\dot{\omega}}{\omega^2}\ll 1$.  In order to investigate whether this  adiabatic condition is violated in some region outside the horizon,
implying the  break down of the WKB approximation, we are interested in the behavior of  $\frac{\dot{\omega}}{\omega^2}$ at different values of $r$. Using \eqref{frequ}, \eqref{omegdot}, let us define
\begin{eqnarray}\label{dot2}
\delta_\omega\equiv \frac{\dot{\omega}}{\omega ^2}=\frac{1}{4M \omega_{0}}V_r^3 (1-V_r^2)\,.
\end{eqnarray}
It can be checked that the  maximum value of $\delta_\omega$, indicating a  maximal violation of the adiabatic condition for the field modes, is found to be
\begin{eqnarray}\label{maxdot2}
\delta_\omega^{\rm max}\simeq\frac{0.18}{4M \omega_{0}}\,.
\end{eqnarray}
This maximum is located at $V_r=\sqrt{{3}/{5}}$, which corresponds to $r=3.3 M$. In FIG. \ref{figure:add3} we plot  $\delta_\omega$  as a function of distance $r$ to exhibit this behavior explicitly. This shows that the maximal break down of the WKB approximation is found almost at the same radius where our previous analysis suggested the core of Hawking radiation to be generated, lending further evidence to our claim.

\section{Discussion}\la{sec:discus}

Our analysis showed that there are several ways to look at the origin of Hawking quanta providing strong hints 
to a long distance origin of the Hawking radiation. In particular, we saw that one can use the difference in the energy density as considered in \eqref{ro} and the one we computed for a freely-falling observer in \eqref{energy density} from the RSET, to make evident the discrepancy between the naive expectation that a thermal flux exists all the way to the horizon and the actual energy density  for a freely-falling observer in the Unruh vacuum.


 We also found  that,  after including the term capturing the effects of non-exact adiabaticity of the  temperature functional, the discrepancy with the energy density obtained from the RSET increased in magnitude. One might think that this is contradictory as the energy density considered in \eqref{energy_mod} closely resembles the energy density computed from the RSET in \cite{Barbado:2016ger}, but this is not the case. 
This can be accounted for by the fact that the quantity computed in  \cite{Barbado:2016ger} is not the same energy density for the freely-falling observer as we computed in \eqref{energy density}. A ``perceived'' stress energy tensor was computed in \cite{Barbado:2016ger}  by subtracting a contribution, as measured by the observer in a local vacuum state defined in its local inertial frame (based on the coordinate $(u,v)$), from the energy density as measured in the vacuum state globally defined based on the null coordinates $(U,V)$.

 The fact that the location of the maximum of this discrepancy happens to coincide with the region of breakdown of the WKB approximation for the test scalar field modes, as we showed in FIG. \ref{figure:add3}, strongly supports an interpretation of our results as evidence for a quantum atmosphere surrounding any evaporating black hole and peaked around $3M$, from where  the Hawking quanta originate.  
 
 The implications of particle creation at finite distance from the black hole horizon can play an important role in the debate on
 long standing problems such as the information loss and the transplanckian issues. Moreover,  this particular region around $3M$ is of special interest also because the maxima of the potential for a spherically symmetric black hole is at $3M$. The quasi normal modes (QNM) are computed using the WKB approximation as a wave scattering at the turning point of the potential barrier \cite{Schutz:1985zz}. The relation between these QNMs and the underlying microstates of a black hole has been speculated many times \cite{Hod:1998vk,Maggiore:2007nq,Corda:2012tz}.
  We believe our results about the region of origin of Hawking radiation can provide some further inputs for such investigations.

Finally, let us comment about extending our analysis to the case of a rotating black hole. In a Kerr spacetime, the definition of the vacuum states must be tailored, as the notion of positive frequency modes would change due to the presence of an ergoregion also outside the horizon and to the consequent superradiance. As shown in \cite{Murata:2006pt}, this can be done and a derivation of Hawking radiation can be performed, also it can be shown that in the near horizon limit a scalar field theory in Kerr spacetime reduces to a 2D effective theory.  It is also possible to compute the components of the renormalized stress energy tensor and define the possible vacuum states consistently \cite{Ottewill:2000qh}. Further, one can introduce a coordinate system adapted to the freely falling observers \cite{Natario:2008ej} in Kerr spacetime. It should then be possible to repeat the above analysis and, using the physical intuition provided by the heuristic argument presented in~\cite{Dey:2017yez}, expect similar conclusions. We leave this for future investigations.   
  
\section*{Acknowledgement}

The authors wish to thank Carlos Barcel\'o, Luis C.~Barbado and Ra\'ul Carballo-Rubio for their useful remarks on the paper.
Research at Perimeter Institute for Theoretical Physics is supported in part by the Government of Canada through NSERC and by the Province of Ontario through MRI.  ZM acknowledges the 
Erwin Schr\"{o}dinger International Institute for Mathematical Physics
(ESI) scientific atmosphere. ZM was partially supported
by the Erwin Schr\"{o}dinger JRF fund.

\bibliographystyle{apsrev-title}
\bibliography{hawking-temp}

\end{document}